# Surface Acoustic Waves Equip Materials with Active Deicing Functionality: Unraveled Deicing Mechanisms and Application to Centimeter Scale Transparent Surfaces

Stefan Jacob[1]*, Shilpi Pandey[1], Jaime Del Moral[2], Atefeh Karimzadeh[1], Jorge Gil-Rostra[2], Agustín R. González-Elipe[2], Ana Borrás[2]*, and Andreas Winkler[1]*

[1] Leibniz IFW Dresden, SAWLab Saxony, Helmholtzstr. 20, 01069 Dresden, Germany.

[2] Nanotechnology on Surfaces and Plasma Lab; Materials Science Institute of Seville; Consejo Superior de Investigaciones Científicas (CSIC), Americo Vespucio 49, 41092, Sevilla (Spain)

* stefan.jacob@ptb.de,  anaisabel.borras@icmse.csic.es, a.winkler@ifw-dresden.de

Migrating active deicing capabilities to transparent materials with low thermal conductivity has a high potential to improve the operations of several seminal industries in the automotive, robotic, energy, and aerospace sectors. However, the development of efficient and environmentally friendly deicing methods for these scenarios is yet in its infancy regarding their compatibility with end-user surfaces at relevant scales and real-world operations. Herein, we approach deicing through nanoscale surface activation enabled by surface acoustic waves (SAWs), allowing efficient on-demand deicing of surface areas spanning several square centimeters covered with thick layers of glace ice. We contemplate SAW-based deicing from a twofold perspective: first, we demonstrate its functionality both with a bulk piezoelectric material (LiNbO$_3$) and a piezo-electric film (ZnO), the latter proving its versatile applicability to a large variety of functional materials with practical importance; second, we gain fundamental knowledge of the mechanisms responsible for efficient deicing using SAWs. In particular, we show that SAW vibrational modes easily transport energy over greater distances outside the electrode areas and efficiently melt large ice aggregates covering the materials' surfaces. In addition, the essential physics of SAW-based deicing is inferred from a carefully designed experimental and numerical study. We support our findings by providing macroscopic camera snapshots captured *in situ* inside a climate chamber during deicing and highly resolved laser-doppler vibrometer scans of the undisturbed wavefields at room temperature. Great care was taken to deposit the interdigital transducers (IDTs) used for SAW excitation only on ice-free areas close to the chip edges, leaving most of the substrate used for deicing unaltered and, as a matter of fact, demonstrating transparent deicing solutions. We expect that the results of this study will have a significant theoretical and practical impact on the development of efficient deicing systems for their use in numerous industries and devices.

## Introduction

The accumulation of ice on surfaces frequently causes system malfunction and failure. Among others, this is a common problem in the robotic, energy, and transportation sectors [1–4], where cameras and sensors are particularly vulnerable to ice buildup. Procedures to deice systems with typical dimensions measuring several square centimeters are researched in both industry and academia. Strategies to prevent or detach ice are typically classified as passive or active, depending on whether they intend to avoid the formation of ice or produce their removal by surface activation. Passive systems often employ icephobic coatings that are applied as cover layers to the substrate's surface during fabrication.[5–7] The coatings modify the contact angle of water and the adhesion stress of ice favorably, enabling water roll-off and ice removal by weight and wind. Although materials with excellent icephobic properties are reported in the scientific literature, so far none of them have shown sufficient resistance and durability for their long-term in-field application.[8] In contrast, active systems typically manipulate the surface *in* situ through an externally triggered activation mechanism (such as an electrical power feed in electrothermal systems[9]). This can make them more flexible compared to passive systems, although restrictions still exist for their application to large surfaces and/or temperature-sensitive devices.

Acoustic waves (AWs) and surface acoustic waves (SAW) are generated on the bulk or surface of piezoelectric crystals and film materials such as quartz, ZnO, AlN, LiNbO$_3$, LiTaO$_3$, etc. by exploitation of the inverse piezoelectric effect. The ample excitation conditions (from kHz to MHz range), material compatibility, and duality between sensing and actuation have prompted their application as critical elements in the development of electronic components, telecommunications, biosensors, and microfluidic and lab-on-a-chip devices.[10] Closed to the topic dealt with herein, low-frequency AWs have been also implemented through ultrasonic transducers for the deicing of aeronautical surfaces as reported in reference.[11] More recently, applications of high-frequency vibroacoustic waves for either the deicing of small ice aggregates or rime ice layers or to prevent icing have been proposed.[12,13] A straightforward way for surface integration of acoustic wave-generating devices consists of fabricating Interdigital transducers (IDTs) on piezoelectric bulk materials or piezoelectric films to produce a dynamic surface activation. Using this strategy, in ref [12] the authors observed that SAWs on substrates measuring several square millimeters delayed ice nucleation in small water droplets due to the "acoustic streaming force inside the droplet, and (the) acousto-thermal effect". They also demonstrated deicing through melting and cracking at the ice-substrate interface in the close vicinity or onto the IDTs. Zeng et al. recorded a minimization of surface ice adhesion due to "electrostatic force and mechanical interlocking, and generating interface heating effect".[14] Very recently, Yang et al. researched the mitigation of rime ice formation and the melting of several hundred μm thick, porous layers of rime ice using SAW in piezoelectric films on aluminum plates measuring a few millimeters.[15] They attributed deicing mainly to acoustic thermal effects caused by the SAW piezoelectric film interaction and streaming inside the evolving liquid phase during melting. These studies [12,14,15] all reported an energy-efficient activation of small surface areas. In recent work, we have also demonstrated energy-efficient deicing and prevention of icing with decreased surface ice adhesion using high-frequency thickness shear acoustic vibrations in larger bulk piezo-electric substrates.[16]

Refs. [12,14,15] have opened the path for the application of SAW-based deicing on metal substrates with high thermal conductivity. These authors have also inferred some physics and understanding of the deicing mechanism applicable to small ice volumes (in the order of microliters) and porous "rime" ice. However, it must be noted that the activation of deicing processes on surfaces (generally referred to as *surface activation*) was only performed on the IDTs and in their vicinity, creating effective deicing areas that measure only tens of SAW wavelengths. Additionally, the IDTs covered large parts of the deiced surfaces, rendering the approach impractical when applied to functional materials (e.g. transparent or nano-coated materials). Thus, to prepare the method for exploitation in real-world scenarios, advances are required in critical aspects, such as the activation of much larger surfaces, the ability to melt larger ice volumes including dense and thicker layers of glace iced, and the compatibility with materials such as thermal insulating substrates. In this paper, we go a step forward in the extension of SAW-based deicing to surfaces with dimensions of at least several centimeters, which constitutes an important advancement in the technology. We demonstrate that SAW can easily propagate energy over several centimeters in distance and that the provided energy is sufficient to drive deicing processes far away from the IDT electrodes. Indeed, we show how SAWs can be excited with compact IDTs close to the chip edges leaving most of the deicing surface unaltered. Importantly, we demonstrate the migration of deicing capabilities to transparent, low thermal conducting substrate materials, which, for the first time, enables SAW-based deicing to be used in optical systems.

As an ideal substrate for Rayleigh-SAW excitation, we employed a piezoelectric 128°YX LiNbO$_3$ crystal substrate of the dimensions 70 mm x 35 mm x 0.5 mm. LiNbO$_3$ is often used in industry and academia as a carrier substrate for SAW with well-understood electromechanics properties.[17] Hence, numerical modeling is possible to infer knowledge about the size of interaction areas between SAW vibrational modes and water-ice or water at different deicing stages. Furthermore, we conduct temperature measurements while applying different excitation modes to explore the effect of advertent heat conduction (caused by ohmic losses) from the IDTs and compare this to the energy transport through SAW surface activation. This variable has not been studied until this time. To demonstrate the deicing capabilities for industrially more relevant scenarios and materials, a commercially available optical grade fused silica substrate (*f*-SiO$_2$) of the dimensions 30 mm x 15 mm x 0.5 mm, layered with a 5 µm piezoelectric highly texturized ZnO film is used. The latter is an interesting application test for real-world scenarios, e.g. glass is used in optical sensor windows that are vulnerable to ice buildup. On both tested materials, large parts of the surface were covered with water ice, extending several hundred SAW wavelengths away from the IDT zones. The devices are optically transparent; and we use compact, high aspect ratio IDTs close to the sides of the chips. The demonstrated surface deicing achieved by SAW-activation in films on ordinary glass-like carrier substrates or inexpensive, commercially available, optically transparent, piezoelectric materials may expedite the industrialization of the research results for a broad range of applications.

## Methods

### Device Design and Fabrication

We designed two SAW devices based on 1) a double-sided polished, single-crystalline LiNbO$_3$ chip substrate (further referred to as *LiNbO$_3$ device*) and 2) an *f*-SiO$_2$ plate, coated with a 5 µm thick piezoelectric ZnO film (further referred to as *ZnO device*). In both devices, the SAW excitation was realized with comb-shaped IDT finger pairs, patterned in regular, quarter-wavelength pitches (**Figure 1**).

When a radio frequency (RF) signal is applied to the IDTs, the oscillating electric potential triggers SAWs on the piezoelectric material. Providing that the excitation frequency matched the resonance of the finger structure, constructive interference leads to the development of a SAW field propagating perpendicular to the IDT fingers. The design operating frequency was estimated from the material-dependent SAW velocity and the desired wavelength. The exact frequency was then determined on the fabricated device using S$_{11}$ measurement techniques. Although this is a standard procedure for SAW devices, the design process differed for both devices due to the more complex mechanical properties of the ZnO – SiO$_2$ combination (compared to the better-understood LiNbO$_3$).

**Self-supported LiNbO$_3$ devices.** We applied elastic surfaces wave theory (based on Stroh[18]) to estimate the propagation velocity (c) of the Rayleigh-type SAW mode as c = 3981 m/s (with material properties from Kovac et al.[17]). The device's wavelength ($\lambda$) was set to 120 µm, resulting in a bulk-thickness larger than four wavelengths, favoring pure Rayleigh-type SAW.[19] The excitation frequency ($f = c/\lambda$) for that setup was estimated to be 33.2 MHz. The device dimensions were defined by a circular four-inch (102 mm) standard wafer size, fitting two chips of the dimensions 35 mm x 70 mm per wafer. The IDT aperture was set to 30 mm, enabling us to activate most of the surface of the device with SAW. We used two IDTs per chip, each located 2 mm from the chip edges. In an iterative procedure, we varied the number of IDT fingers to optimize the electromechanical coupling at the design excitation frequency. Good coupling was found for ten finger pairs, which we used for the devices throughout the article. The total area allocated for IDTs and connection pads on the chip was less than 10 % of the total surface area (including the total area spanning from the chip edge to the furthermost IDT finger).

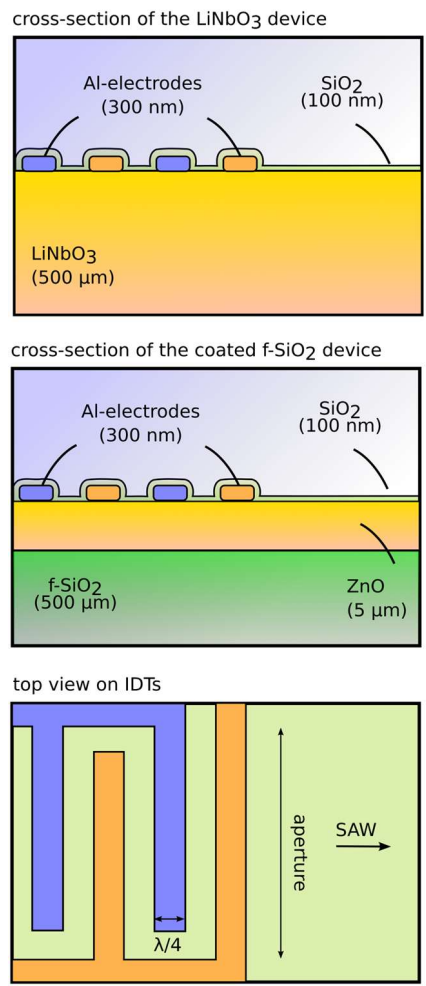

*Figure 1: Sketch of the IDT configurations for LiNBO$_3$ and ZnO devices. From top to bottom: cross-section of LiNbO$_3$ bulk material with aluminum electrodes, topped with a protective layer of SiO$_2$; cross-section of ZnO coated fused quartz (f-SiO$_2$) with aluminum electrodes, topped with a protective layer of SiO$_2$; top-view on comb-shaped, reciprocally polarized electrodes with quarter-wavelength pitch ($\lambda/4$).*

LiNbO$_3$ 128°YX wafers (4", double-side polished, black, obtained from Hangzhou Freqcontrol Electronic Technology Ltd., China) were selected as substrates, due to their well-known material properties, optical transparency, relatively low thermal conductivity, and high coupling coefficient for Rayleigh-SAW excitation. Interdigital transducer electrodes were produced of subsequent layers of Ti (5 nm) and Al (295 nm), and prepared via laser-lithography (Heidelberg Instruments MLA 100), conventional electron-beam evaporation (Creamet 350 CI 6, CreaVac GmbH) and lift-off technique. The whole wafer surface was coated with 100nm $SiO_2$ by reactive RF magnetron-sputtering from a $SiO_2$ target, mainly to protect the Al IDTs from electro-corrosion. Pad opening was realized via laser-lithography and reactive ion etching.

**ZnO devices.** The design of the ZnO device was restrained by the narrow range of possible SAW wavelengths with good electromechanical coupling for a given ZnO film thickness. Dimensionless thicknesses of the film ($\frac{h}{\lambda}$) larger than 0.25 have been reported (Tomar et al. [20]) to achieve the best electromechanical coupling between IDTs and the substrate. Therefore, we designed our IDTs for a wavelength of 20 μm on a 5 μm thick ZnO layer ($\frac{h}{\lambda} = 0.25$). The excitation frequency was estimated to be approximately 135 MHz (with a wave velocity of 2700 m/s depicted in Kovac et al.[17]). The dimensions of the substrate (30 mm x 15 mm) were dictated by the fabrication reactor applied for the ZnO film but can be straightforwardly scalable to larger areas. We used a single pair of IDTs with an aperture of 20 mm; the IDTs were located 2 mm from the edge of the chip. In an iterative procedure, we varied the number of IDT fingers to optimize the electromechanical coupling at the design excitation frequency. Good coupling was found for 45 finger pairs, which we used from now on. The total area allocated for IDTs and connection pads on the chip was less than 25 % of the total surface area (including the total area spanning from the chip edge to the furthermost IDT finger).

Optical grade fused $SiO_2$ (f-$SiO_2$) (obtained from Vidrasa, Spain) was selected as substrate, due to their excellent optical properties and compatibility with various industrial processes, especially in optical windows. Polycrystalline (002) textured ZnO films of approx. 5 μm thickness were deposited on the fused silica substrates by pulsed DC reactive magnetron sputtering at ambient temperature. The power source was operated at a power of 150 W. A mixture of Ar (15 sscm) and $O_2$ (15 sscm) at a total pressure of 5.0 $10^{-3}$ mbar was dosed in the chamber to ignite the plasma. The deposition was controlled with a quartz crystal monitor. IDTs and $SiO_2$ were processed as described in the previous section for the Self-supported $LiNbO_3$ devices.

**Figure S1** in the supporting information gathers the UV-VIS-NIR transmittance spectra for both the $LiNbO_3$ plate and the ZnO film showing high transparency. The Spectra were acquired in a PerkinElmer Lambda 750 UV/vis/NIR model.

### Device and Performance Testing

The electromechanical coupling and general function of the devices were tested after fabrication using a network vector analyzer (Keysight ENA E5080B, Keysight Technologies, CA, USA), measuring the electro-mechanical coupling ($S_{11}$ vs. frequency) of the electrodes at room temperature (i.e. 25 °C).

The amplitude and phase of the surface normal displacement at the design frequencies were measured with a laser doppler vibrometer (Polytec UHF-120, Polytec GmbH, Germany). Therefore, the IDTs were activated with a weak RF signal (i.e. less than 3 V peak-to-peak voltage) at the operating frequency. Parts of the surface were then sampled on a regular grid, resolving the vibrations with at least ten points per wavelength in the main SAW propagation direction and two points per wavelength perpendicular to the propagation direction. The data presented in the Results section consists of 25,000 points; the measurement took approximately two days for each of the samples. During this time, the focus point of the vibrometer was recalibrated every 500 measurement points using an automized, optic-based routine. In addition, line scans of the normal displacement on the centerline of the device (in the main SAW propagation direction) were recorded. A spatial Fourier Transform analysis was conducted to extract the wave numbers of the dominant vibrational modes. The wave numbers were compared to the theoretical values (using the expected SAW speed and the excitation frequency) to prove the dominance of SAW modes in the deicing experiments.

Furthermore, the roughness of the surface for the ZnO devices was measured using atomic force microscopy (AFM) (Dimension Icon, Bruker, MA, USA) in the tapping mode in the scanning area of 20 x 20 μm². The scanning area was considered large enough to eliminate the length scale effect on the RMS roughness value.

### Deicing testing

Deicing was tested in a climate chamber (Binder MKF-56, BINDER GmbH, Germany). The test setup contained a chip holder, a LED light source, a microscopic camera, a coaxial SMA cable, and a portable multifunctional rf signal source (BelektroniG BSG, BelektroniG GmbH, Germany).

A custom chip holder setup was made from flat PEEK polymer with two adjustable PCB holders. The customized 50 Ω matched printed circuit boards (PCB, manufactured by Würth Electronic GmbH & Co. KG., Germany) with gold-coated spring pins contacted directly to the Al pads on the chip surface.

A typical deicing test consisted of the following steps:

1. The deicing chip was placed in the climate chamber (at 20 °C). The $S_{11}$ vs. frequency was measured to guarantee the correct function of the device. During the experiments, $S_{11}$ vs. frequency was sampled using the portable rs signal source. A typical $S_{11}$ spectrum can be seen in **Figure 2**. SAW excitation was realized at the frequency with the lowest $S_{11}$ value. It can be seen that the range of low $S_{11}$ extends through a frequency range of several hundreds of kHz, allowing for a very robust excitation.
2. We dripped distilled water to the device's surface using a syringe with volumes of 3 mL and 1 mL onto the LiNbO$_3$ and ZnO devices, respectively, to cover most of the substrate's surfaces outside the IDT region. The distance between water and IDTs was 10 mm and 2 mm in LiNbO$_3$ and ZnO devices, respectively, to ensure no influence of water loading on the SAW excitation.
3. The temperature in the climate chamber was gradually reduced to the target temperature (e.g., -15 ºC) for a time of 15 min. The temperature was then held constant until the water froze. The cooling process was slow enough to get a homogeneous cooling of both chamber walls and device, thus precluding any preferential condensation of the water humidity in the air of the chamber at 20ºC (i.e. before the cooling process) on the sample surface. These cooling conditions rendered a rather transparent layer of glace ice and a dry atmosphere with a negligible vapor pressure of water.
4. The SAW excitation was enabled, to initiate the deicing process. The deicing was monitored using a macroscopic camera. During the experiment, the power supply was restricted by the portable rf signal generator to a maximum of about 7 W (depending on the frequency and its corresponding $S_{11}$ parameter). Therefore, a typical deicing experiment took between 5 to 10 minutes for either of the devices.

Videos of the deicing experiments are included in the supplementary material. **Videos S1** and **S2** show deicing experiments on the LiNbO$_3$ and ZnO devices, respectively. For **video S3**, ice containing red-colored polystyrene particles type PS-FR-Fi252, micro-particles GMbH, Germany, concentration approx. 1%) was melted, allowing for visualization of the physical effects during the different melting processes.

**Numerical Computations**

We carried out a numerical study to understand the mechanical interaction of the SAW with the water and the ice. We modeled a LiNbO$_3$ plate with IDTs (but without a holder and circuit board) using a predefined set of fully coupled electrodynamic and mechanic equations in the commercial finite element (FE) solver COMSOL Multiphysics version 6.0 (similar to the numerical setup described in Fakhfouri et al.[21] ). Since the Rayleigh-type SAW on 128° X-Y LiNbO$_3$ is mainly polarized in the median plane[22], the system was modeled in two dimensions with a generalized plane–stress assumption. The substrate was described with an anisotropic, linear-elastic, and piezoelectric material model using the stiffness, piezoelectric, and permittivity tensors for LiNbO$_3$ (depicted in Kovacs et al.[17]), which were rotated to reflect the crystal cut used in the experiments. Furthermore, we added domains for the water and ice (as a drop with the shape of an elliptic arc) and the surrounding air. The fluids were modeled with a quiescent, linearized form of the Navier-Stokes equations (termed "Thermoviscous Acoustics" in COMSOL Multiphysics), using the material properties from the internal material library of the software. We used a linear-elastic mechanical model for the ice, applying the poly-crystalline (isotropic) material properties from Victor and Whitworth[23].

The geometry was discretized with an unstructured triangular mesh, resolving the acoustic modes with at least six elements per wavelength (for $c_{substrate} = 3981 m/s$ [17], $c_{water} = 1400\ m/s$, $c_{ice} = 1850 m/s$ [23], $c_{air} = 340\ m/s$ at $f = 32.5\ MHz$ ). In addition, the thermoviscous boundary layer at all fluid-solid boundaries was resolved with a stretched boundary layer mesh. This resulted in a numerical grid with approximately three million elements (or approx. 18.3 million degrees of freedom, depending on the exact composition of the ice and water phase). Increasing the mesh density further did not noticeably change the results.

The IDTs were modeled as one-dimensional, massless, and mechanically soft patches to the surface of the bulk material (similar to Fakhfouri et al.[21]), mimicking the pattern of the IDTs used in the experiment. The SAW was triggered by applying a harmonic electric potential to the IDTs. At the edges of the computational domain, the waves were absorbed with perfectly matched layers (PML, predefined in the commercial solver). The stationary acoustic field for a single excitation frequency (i.e. 32.5 MHz, as used in the experiments) was computed by assuming time-harmonic solutions.

## Results and Discussion

The three most important findings of our study can be summarized as follows:

1. The surface of piezoelectric bulk material and piezoelectric films can be activated for deicing using high aspect-ratio IDTs located close to the edges of the chip, only covering small portions of the substrate material's surface, which in addition can be a transparent low heat conductor. The rf excitation of SAW-based chips is simple and robust against temperature changes and water-ice loads on the substrate, allowing an excitation at fixed frequencies. This is advantageous over similar approaches employing more temperature-sensitive acoustic wave modes.[16]

2. SAWs can transport energy over distances of several hundred wavelengths (usually several centimeters) until they interact with water or water-ice interfaces. During the interaction, the acoustic energy in the SAW provides the excitation energy required to drive ice melting or prevent ice formation processes working under active anti-icing operation. The physics involved in the deicing is specific to the partition distribution of water and ice phases.

3. Energy is provided mainly by the SAW leaking into the water/water-ice and ice-water/substrate interfaces, and not by heat conduction in the bulk material.

In the following sections, we discuss those findings in detail.

**SAW excitation and propagation.** The $S_{11}$ vs. frequency spectra are presented in **Figure 2**, proving electromechanical coupling at frequencies between 33 MHz and 35 MHz (LiNbO$_3$ device) and 133 and 135 MHz (5 µm ZnO device). The signal modulation observed for the LiNbO$_3$ device results from SAW reflection at the chip edges and opposing IDTs. The higher SAW attenuation in the ZnO device almost completely suppresses this effect. Temperature coefficient of frequency (TCF) of -86.3 ppm/°C for 128° XY LiNbO$_3$ (from Ward [24]) and -10 ppm/°C for 5 µm ZnO on SiO$_2$ (from Tomar [20]) led to an estimated maximum shift of the SAW frequency of 0.5 % (<0.2 MHz) and 0.07 % (<0.1 MHz), respectively, for temperature ranges relevant for deicing applications (i.e. -40 °C to 25 °C). This was confirmed by the $S_{11}$ measurements at 25 °C and -25 °C in **Figure 2**, showing a mostly constant frequency range for the best electromechanical coupling. Compared to the high temperature sensitivity of other wave modes (e.g. in Ref. [16]), deicing with SAW, therefore, enables a much easier signaling approach with simpler electronics. This finding agrees well with those in Yang et al.[15]

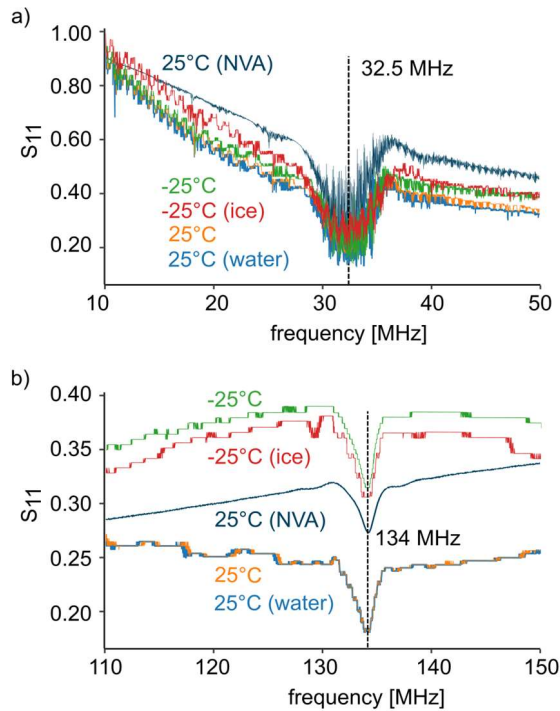

*Figure 2: $S_{11}$ over frequency for different temperatures and mechanical loadings, for a) the LiNbO$_3$ device and b) the ZnO device, measured with a network vector analyzer (VNA). The dips indicate electromechanical coupling between the substrate and the IDTs. The frequency of best coupling does not noticeably shift for the different conditions.*

Furthermore, we present $S_{11}$ data for water-loaded and glace ice-loaded surfaces (>50 % of the surface was wetted); no significant difference was observed between the coupling frequencies of the free substrate surface and the surface covered with water or ice. The temperature- and loading-related frequency shifts were much lower than the bandwidth of the electromechanical coupling. Therefore, the frequency shift of the electromechanical coupling was negligible in the experiments; this led to a simplified SAW excitation at static frequencies. This finding also highlights the robust mode of operation for the SAW-based deicing devices. It should be noted that the broad-band values of $S_{11}$ are lower than one, which indicates losses (e.g. capacitive and ohmic) during the acoustic wave excitation.

**Figure 3** shows the surface normal displacement (snapshot in time) from LDV measurements for both devices at their coupling frequencies (32.5 MHz for the LiNbO$_3$ device and 134 MHz for the ZnO device). It proves the clear presence of wavefronts with a wavelength corresponding to the pitch of finger electrodes of the same electric potential (**Figure 3** a)), suggesting a dominant excitation of Rayleigh-type SAW modes. It should be noted that the voltage of the RF excitation for the LDV measurement was adjusted individually for each device to achieve a good signal-to-noise ratio. Hence, although the activation of the surface is clearly demonstrated, the amplitude values of the wavefronts are not informative. The wavefronts were found to be uniform over the entire aperture (for reasons of clarity, this is not plotted in **Figure 3**), suggesting a spanwise uniform activation of the surfaces. In addition, line scans along the center line (in the direction of wave propagation) are presented (**Figure 3** b)). The consistent wave-like pattern of the oscillation was detectable over the entire line scan for both devices, proving far-field propagation of the waves up to several hundred wavelengths distance from the IDTs.

The vibration of the ZnO device was characterized by decay of approximately 1.5 dB / 100 λ, which most likely is due to the roughness and imperfections in the polycrystalline ZnO film (as argued by Wu et al.[25] for ZnO layers with similar propagation losses). The average roughness (RMS) of the ZnO device measured with the AFM method was 9.96 nm. In contrast, the LiNbO$_3$ device (double-sided polished single-crystalline substrate, roughness less than 0.5 nm) showed no visible propagation loss. We observed a weak overtone (visible as a modulation in the vibration signal), indicating the presence of at least one weak inadvertent wave mode. This was confirmed by using spectral plots, which revealed a strong tone for both devices, with some weaker tones for the LiNbO$_3$ device (**Figure 3** c)). Note that the propagation velocities of the main tones correspond to the velocity of the fundamental Rayleigh modes on the substrates, showing that the observed vibrations originate from the intended SAW modes. The spectral plots were computed using a spatial Fourier Transform of the vibrometer scan.

**Deicing mechanisms.** The deicing experiments are presented for the LiNbO$_3$ device and the ZnO device in **Figure 4**. Corresponding videos can be found in the supplementary material (**Videos S1** and **S2**). The settings of the deicing experiment are summarized in **Table 1**.

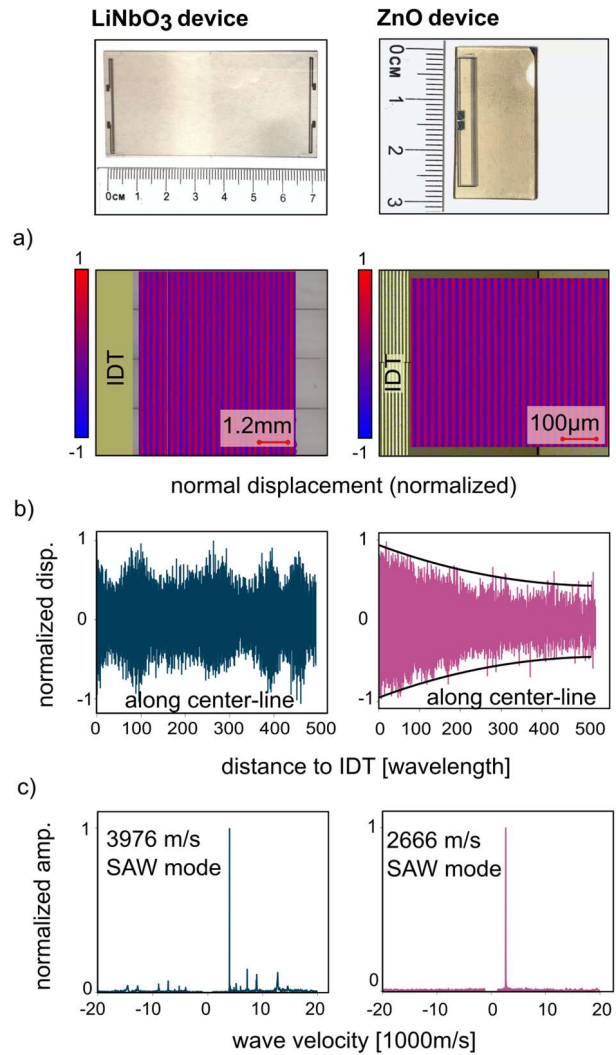

*Figure 3: Surface normal vibration at design frequency for LiNbO$_3$ (Left) and ZnO devices (Right), measured with a Laser - Doppler Vibrometer (LDV). Top: Photo of the experimental device and bottom: a) visualization of a fully resolved wave field from LDV measurements (normal component); b) displacement (real part) along a center line of the chip from LDV measurements (normal component); and c) spectral analysis of the normal displacement along the center line, suggesting the presents of strong directional SAW components. The values in the vertical axes of the line plots are normalized to the largest value.*

We found similar deicing principles for both substrates, i.e. we discovered that the deicing can be broken down into three temporal processes. First, the deicing was initiated by the formation of a thin water film along the ice edges directed towards the activated IDTs (*initial deicing process*). This formation appeared with a delay of a few seconds (5-20 s in our experiments) after the surface activation. The water film was observed along the entire front of the ice, directed to the IDTs. Second, the ice gradually melts (*melting process*). Circular reaming of small air bubbles was observed in the water. We also observed partly dewetted substrate areas. Third, after the ice was completely melted, the water stayed liquid as long as the SAW was activated (*active anti-icing process*), whereas the remaining water froze again after deactivating the SAW.

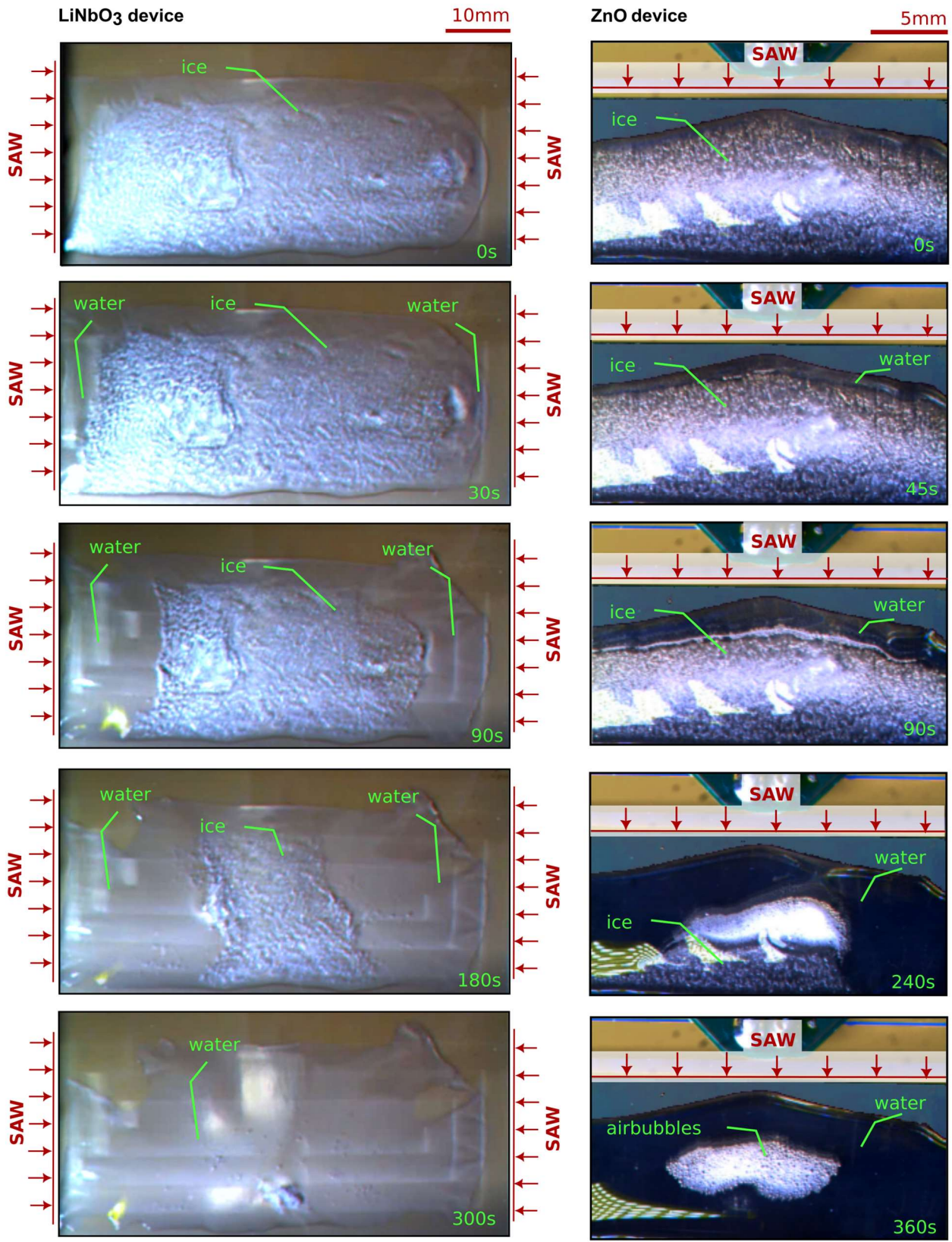

*Figure 4: Snapshots of the SAW induced deicing process on large substrates of LiNbO$_3$ (left) and ZnO on f-SiO$_2$ (right). A wetted area of 18 cm$^2$ / 3 cm$^2$ covered with glace ice was deiced. The applied SAW wavelength was 120 µm / 20 µm, resulting in an excitation frequency of 32.5 MHz / 134 MHz. The camera is placed at the top of the substrate and normal to the surface.*

*Table 1: Settled conditions during the deicing tests.*

|  | Wetted Area | Air Temperature | Frequency | Power | Activation Time |
|---|---|---|---|---|---|
| LiNbO$_3$ device two IDTs | 6.5 cm x 3.0 cm | -10 °C | 32.5 MHz | 2 x 3.5 W | 5 min |
| ZnO device One IDT | 2.0 cm x 1.0 cm | -15 °C | 134 MHz | 1 x 7.5 W | 6 min |

The interaction of the SAW on LiNbO$_3$ with water and ice during the different deicing processes is shown in **Figure 5**. In this figure, we present the numerical simulations of the multi-physics model computed in the commercial FEM solver COMSOL Multi-physics. The results were used to depict the effective area where the SAW interacts with ice/water (which we denote as the *interaction zone*) in the case of large water/ice accumulations. The size of the interaction zone was measured based on the exponential ($1/e$) decay distance of the SAW. **Figure 5** a) shows a reference case of the free substrate hosting a SAW that propagates from the left side of the plot to the right side. For **Figure 5** b) the same substrate was loaded with ice. Here we see that the SAW interacts with the ice during the *initial deicing process*, and the amplitude of the SAW displacement vector at the LiNbO$_3$ – ice interface decays sharply along the ice-substrate interface. This restricts the SAW – ice interaction zone to a distance of approximately five wavelengths. Unlike the hypothesis in Ref. [15] for porous micrometer rime ice layers containing a substantial volume of air pockets, we did not observe strong SAW reflections at the ice-substrate interface, very likely because of the different nature of the ice (rime and glace) formed in each case. In **Figure 5** c) the substrate appears loaded with well-defined water and ice adjacent fronts, mimicking a snapshot of how the glace ice melts during the *melting process* (compare to **Figure 4**). The SAW decays slower at the water–substrate interface than at the ice-substrate interface, enabling the interaction zone to extend further along the interface (approximately seven wavelengths in the computed case). For **Figure 5** d) the substrate was loaded with pure water. In this phase, we observed that the interaction zone reaches a maximum length of approximately ten wavelengths, which is also the theoretical value reported in the literature for fluid-solid interfaces.[26]

The experimental study and the numerical results imply that the mechanisms of substrate–load interaction change during the deicing process. The different mechanisms that lead to deice surfaces are presented as a sketch in **Figure 6**. The short interaction zone in the *initial deicing process* (**Figure 6** a)) produces regions of large mechanical stresses in the vicinity of the ice-substrate interface at the edge of the ice phase (directed towards the IDTs). This mechanism was also considered recently by Yang et al. [12] studying the SAW effect on small and compact glace ice aggregates, (termed by the authors as *nanoscale "earthquakes"* to stress the small and local character of the interaction) on vertically oriented substrates. They attributed the melting during this stage mainly to crack development and crack growth around microscopic air inclusions at the interface and to thermo-acoustic effects of the leaking SAW. In Ref. [15], after long activation, the authors monitored the collapse of ice structures during an "initial stage" in thin layers of porous rime ice, eventually leading to thin water films with unmelted ice aggregates. This results from a relatively slow process due to the initial high acoustic reflectance by the air pockets present in the ice structure. Although the magnification of the camera used in our experiments did not allow us to observe sub-millimeter melting and crack growth, the development of the thin water film at the confined locations of the SAW – ice interactions that were observed during the initial phase supports this assumption. The presence of such a thin water film without ice aggregates would be also consistent with the finding in our previous article dealing with deicing by "bulk" acoustic waves.[16]

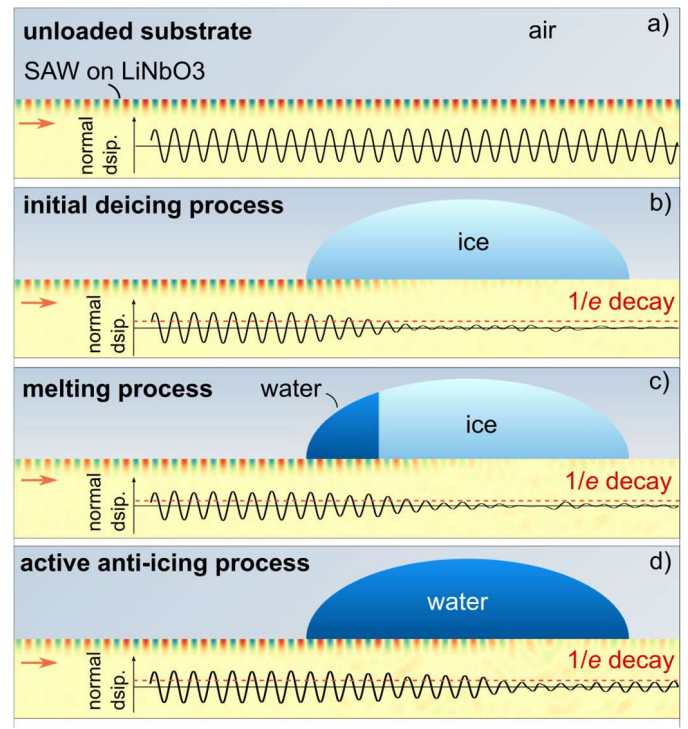

*Figure 5: SAW interaction with water and glace ice for the different deicing processes. a) reference model without water/ice. b) pure ice, indicating a small SAW interaction zone (5 λ), c) water and ice, indicating an extended interaction zone (7 λ), d) pure water, indicating a large interaction zone (10 λ). Computed with COMSOL Multi-physics.*

During the *melting process* (**Figure 6** b)), the SAW efficiently leaks longitudinal pressure waves under the Rayleigh angle into the evolving water film, where they were dissipated by the viscosity of the fluid, creating heat and streaming (visible in **Video S3**).[27] SAW-induced heat leads to a significant temperature increase in water droplets (as demonstrated by Shilton et al.[28]). Heat is not only created in the bulk of the fluid but also in the thin thermo-viscous boundary layers at fluid-solid interfaces (as shown by Muller and Bruus [29]). In our case, these mechanisms would lead to strong heat currents along the well-defined ice-water interface where melting occurs. Additionally, the mechanical interaction at the substrate-ice interface continues (at least until a critical depth of the waterfront of about ten wavelengths is reached), which also provides a progressive interfacial melting resulting in the development of thin water inclusions between the substrate and ice. This effect was clearly visible during the experiments (shown in **Videos S1** and **S3**).

The temperature difference within the fluid causes convection, which subserves internal flow and therefore an efficient energy transport. In addition, microfluidic streaming lobes driven by SAW-induced pressure gradients steer the fluid.[30] Therefore, the melting continued, even though the SAW was not directly interacting with the ice once the water phase exceeded ten wavelengths. Since the process was driven by a constant flux of energy from the SAW into the water, the melting decelerated visibly when the surface of the fluid-phase increased and the water conducted more heat to the surrounding air and the substrate. The heat flux ($\dot{q}$) from the water to the environment can be expressed as $\dot{q} = A\, h\, \Delta T$, where $A$ is the contact surface between water and the environment, $\Delta T$ is the temperature difference, and $h$ is the specific heat transfer coefficient. The specific heat transfer coefficient of air is between 10 W/m²K and 600 W/m²K, depending on the air movement.[31] This leads to a minimum heat flux from the melted water to the surrounding air in the order of several hundred milli-Watts for the LiNbO$_3$ devices tested in this paper. However, the substrate itself could act as a heat sink. The remaining (*effective*) heat in the water promotes the melting process. The energy required was mainly the latent heat, which is 334 J/g for ice[23], leading to the typical melting time in the order of several minutes observed for the fully loaded devices tested in this paper (with loads between 1 g and 3 g of ice and input power of less than 10 W as used in the experiments).

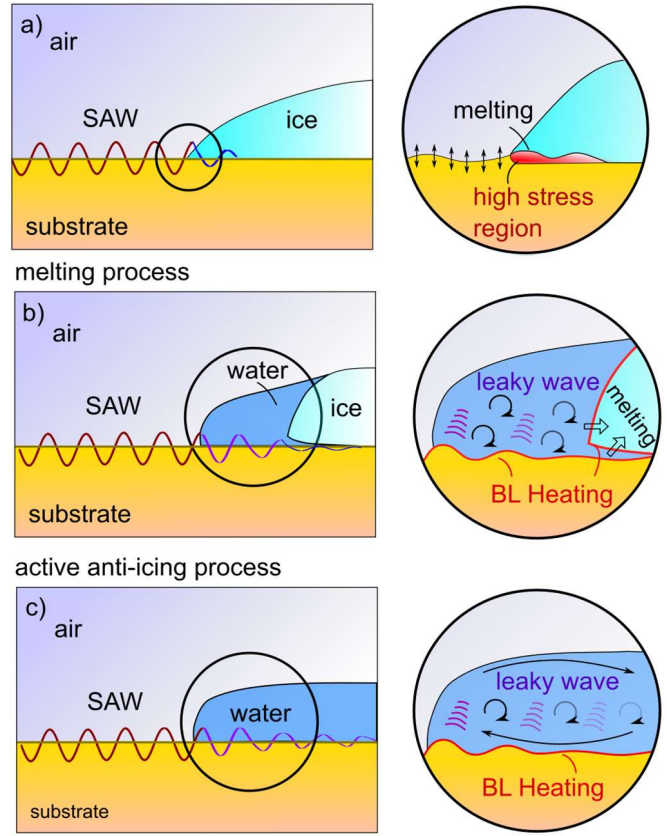

*Figure 6: Sketch of the SAW-induced melting process of glace ice aggregates.* a) initial deicing process: the SAW interacts with the ice, creating regions of high stresses, initiating micro-cracks in the ice and local melting. A thin water film forms in the high-stress region. b) melting process: the SAW efficiently leaks into the water-film and energy is exchanged due to viscosity of the water in the bulk and at the water-ice and water-substrate thermo-viscous boundary layers. The temperature gradients in the fluid initiate streaming and therefore efficient steering. The ice melts at the well-defined interface with the water. c) active anti-icing process: the ice is completely melted. The SAW leaks into the water, creating a constant flux of energy. Large streaming lobes form due to the temperature differences between water, substrate and surrounding air, efficiently mixing the water and preventing freezing.

After melting the whole ice, the water stayed liquid as long as the energy flux was retained and the environmental conditions did not change (*active anti-icing process*, **Figure 6** c)). In our experiments, we observed large streaming lobes (visible through the entrained air bubbles (**Videos S1 - S2**) and particles (**Video S3**). This flow transported heat from the area that was interacting with the SAW to the innermost volume of the fluid (providing sufficient SAW power as demonstrated by Beyssen[32]), keeping the droplet in liquid form. The mechanisms of liquid steering through streaming and convection were of the same micro-fluidic nature as observed during the *melting process* (discussed above).

Those findings are in partial agreement with previous fundamental studies on SAW deicing, which investigate two special cases of ice, i.e., small areas of glace ice[12] and very thin layers of porous rime-ice,[15] formed on or very close to the IDTs. Besides the different characteristics of ice considered in each case, another significant difference between these works and our experimental conditions resides in that we use here ice formed on poorly thermal conductive substrates, which preclude efficient heat conduction from the IDT zone and their surroundings up to the melting zone. For small droplet-size glace ice aggregates, the SAW interaction zone during the initial melting process is in the order of the droplet diameter, resulting in fast and complete melting of the ice-substrate interface during the *initial melting process.* In the case of thin layers of glace ice (with thicknesses of sub-wavelength scale), only a small portion of the energy carried by the SAW is leaking into the fluid, resulting in extended interaction zones and prolonged interfacial substrate-ice interaction.

**Discarding a dominant heating mechanism.** Previous research has shown that the temperature in the IDT region can increase significantly during the excitation stage.[33] This temperature increase has been attributed mainly to ohmic losses in the IDTs. With the experiments gathered in this section, we aim to verify whether the melting of the ice is also driven by such heat released from the IDTs (which is not part of the SAW field). In our study, the effect of heat flux from the IDTs on the melting process was assessed through specially designed temperature measurements during a deicing cycle. Thus, we must stress again that the transparent substrates and piezoelectric films (see **Figure S1**) used in the herein-reported experiments have a relatively low thermal conductivity, much lower than that of metal substrates and that the ice has been formed avoiding the IDTs area. We mounted a type K thermocouple at the back side of a $LiNbO_3$ device (at the center of the IDT region) using silver ink. The thermocouple probe was isolated from the surrounding air using a 2 mm thick isolation tape. No significant influence of the probe on the wave excitation was observed (assessed through $S_{11}$ vs. frequency measurements with and without the probe). The surface of the substrate was covered with 3 ml of distilled water 10 mm away from the IDTs, which was cooled in the climate chamber until freezing occurred (the climate chamber's temperature was set to -20 °C, and freezing occurred at a substrate temperature of 10 °C). We performed two tests: under condition i) with excitation (3 W) at an off-design frequency (42 MHz), where no significant acoustic wave could propagate and only ohmic losses in the IDTs occurred; and under condition ii) with excitation (3 W) at a design frequency (32.5 MHz), where the Rayleigh-type SAW could propagate and ohmic losses in the IDTs occurred. The results are presented in **Figure 7**. The most important results are the points a), b), c), and d) in the temperature-over-time plot and the corresponding snapshots of the ice front directed towards the IDTs.

We observed sharp increases in the temperature when excitation was enabled, followed by sharp drops in the temperature when the excitation was disabled. For both excitation conditions, we observed a similar time dependence for the temperature evolution, though with higher incremental values for condition i) than for condition ii) (i.e., ΔT is approx. 40 ºC and 30 ºC, respectively). The difference in the measured temperature reflects that the losses increase with frequency, as documented in **Figure 2**. We would like to emphasize that the ice did not undergo visible changes during 5 minutes of the condition i) excitation (snapshots a) to c)), but extended melting was observed after 5 minutes of condition ii) excitation (snapshot d)). These results evidence a negligible effect of the ohmic heat released by the IDTs compared to the effect of SAW on the deicing of large surfaces outside the IDT area for low thermal conductive materials. In other words, the heat flux at the ice edge separated 10 mm from the IDT is negligible (note that the thermocouple notices the release of heat because it is placed at the back of the device, just at 0.5 mm from the IDT heat resistance source). We would like to note, that this finding does not apply to cases where either the ice covers the IDTs or devices are built onto metallic substrates (e.g. as reported in Refs.[12,15]). In this case, the ice and liquid phases might efficiently absorb the heat that is released by the IDTs acting as electrical resistance heaters, or partially "electrothermal" deicing devices. Those cases were not considered in our study.

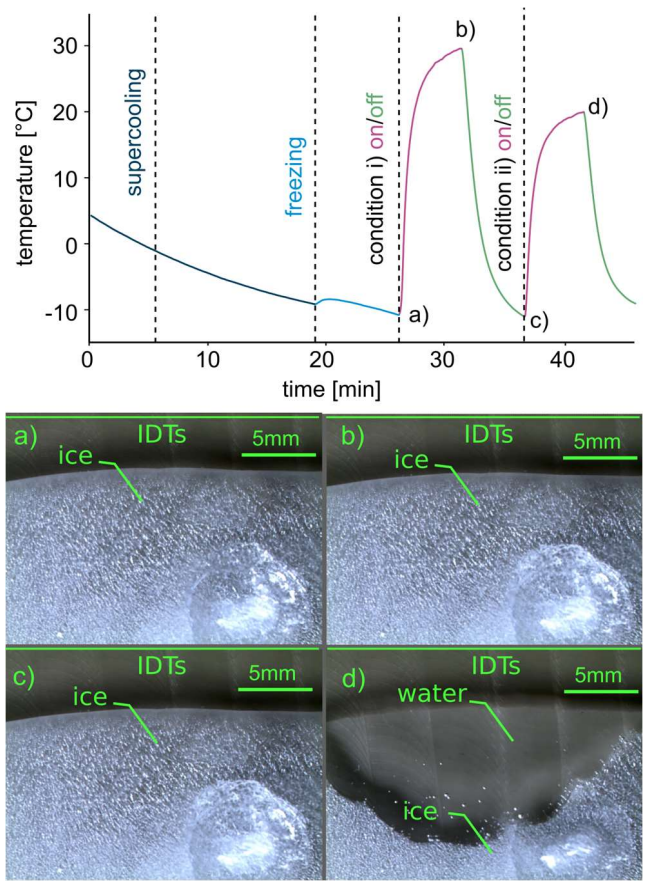

*Figure 7: Assessment of the effect of IDT-heating on the melting of the ice. Top) Temperature was measured with a thermocouple mounted on the backside of the substrate for condition i) under excitation at an off-design frequency (42 MHz), where only omic losses occurred (and no SAW); and condition ii) under excitation at the design frequency (32.5 MHz), where Rayleigh-type SAW modes could propagate. Bottom) The snapshots a)-d) demonstrate that the melting due to IDT heating is insignificant for the ice accreted out of the IDTs areas. The ice was located approx. 10 mm from the IDTs (the line in the snapshots only indicates the orientation of the IDTs, not their location).*

## Conclusions

In this article, we researched SAW-based deicing of material surfaces with dimensions measuring tens of square centimeters. In addition to the scientific value of a better understanding of the deicing mechanisms, the results have high practical value because the dimensions of the tested devices (up to 7 cm and 3 cm in the case of $LiNbO_3$ and ZnO film correspondently) are relevant for many industries that suffer from ice buildup. Most importantly, we showed that SAW can easily transport energy over larger distances to large areas of glace ice, where the energy efficiently drives local melting processes. Although the actual interaction zone is smaller than 10 wavelengths, acoustofluidic effects trigger processes that can melt iced domains covering much larger dimensions. Those effects differ, depending on the partition distribution between ice and water, and were identified in our study based on macroscopic images of the deicing process.

The key findings of the paper regarding the efficient deicing are valid for both of the tested carrier substrates, although the ZnO film on optical grade fused $SiO_2$ showed some SAW attenuation at greater distances. The attenuation was likely caused by the insufficient thickness of the film, its polycrystallinity, and the roughness of the film's surface, the latter parameter determined by the magnetron sputtering procedure utilized to prepare the films. Future research on industrially relevant materials should focus on piezoelectric films and surface engineering procedures leading to surface terminations compatible with SAW activation. In addition to the wave attenuation, we encountered ohmic losses in the IDTs during surface activation. Some of the losses are attributed to the specific design of the electrodes that connected the device to the RF signal generator, which should be also optimized when used in real-world scenarios. Nevertheless, we demonstrated that in poor heat conducting substrates the heat created by those loss mechanisms at the IDT zone does not contribute to the deicing observed in the experiments and that the identified mechanisms are a result of the interaction of water, ice, and the SAW vibrational mode.

## Acknowledgments

The project leading to this article has received funding from the EU H2020 program under grant agreement 899352 (FETOPEN-01-2018-2019-2020 - SOUNDofICE). We are grateful for the received funding that allowed us to conduct the research presented in the paper. Furthermore, we thank Steve Wohlrab for helping us prepare the LiNbO3 and ZnO devices. We also thank Armaghan Fakhfouri for supporting the numerical computations.

# SUPPORTING INFORMATION SECTIONS

# Surface Acoustic Waves Equip Materials with Active Deicing Functionality: Unraveled Deicing Mechanisms and Application to Centimeter Scale Transparent Surfaces

Stefan Jacob[1]*, Shilpi Pandey[1], Jaime Del Moral[2], Atefeh Karimzadeh[1], Jorge Gil-Rostra[2], Agustín R. González-Elipe[2], Ana Borrás[2]*, and Andreas Winkler[1]*

[1] Leibniz IFW Dresden, SAWLab Saxony, Helmholtzstr. 20, 01069 Dresden, Germany.

[2] Nanotechnology on Surfaces and Plasma Lab; Materials Science Institute of Seville; Consejo Superior de Investigaciones Científicas (CSIC), Americo Vespucio 49, 41092, Sevilla (Spain)

* stefan.jacob@ptb.de, anaisabel.borras@icmse.csic.es, a.winkler@ifw-dresden.de

**Video S1.** Video showing a deicing experiment on the LiNbO$_3$. Available from the author.

**Video S2.** Video showing a deicing experiment on the ZnO devices. Available from the author.

**Video S3.** Video showing a deicing experiment on the LiNbO3 for ice containing red-colored polystyrene particles. Available from the author.

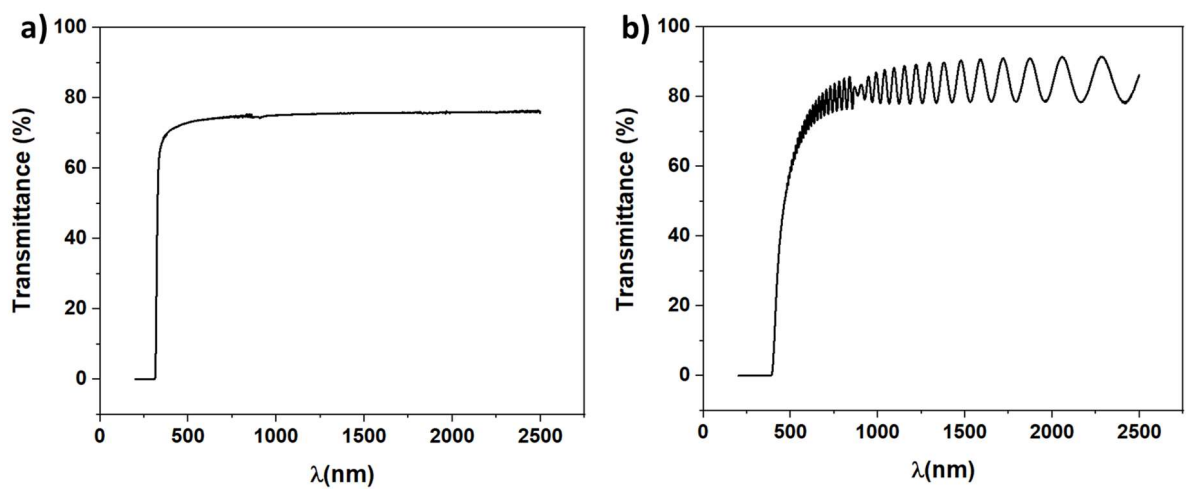

**Figure S1.** UV-VIS-NIR transmittance spectra for LiNbO3 (a) and a 5 μm ZnO film deposited on optical grade fused silica (b).